\documentclass[aps, amsmath, twocolumn, amssymb,superscriptaddress, prd]{revtex4-2}

\usepackage{graphicx}
\usepackage{dcolumn}
\usepackage{bm}
\usepackage{tikz}
\usepackage{quantikz}
\usepackage{physics}

\usepackage[hidelinks, breaklinks, colorlinks=true, linkcolor=blue]{hyperref}
\usepackage[capitalize]{cleveref}
\usepackage{siunitx}

\usepackage{booktabs}
\usepackage{multirow}
\usepackage[table]{xcolor}

\usepackage{moresize}
\usepackage{array}

\crefname{equation}{Eq.}{Eqs.}
\Crefname{equation}{Eq.}{Eqs.}

\crefname{figure}{Fig.}{Figs.}
\Crefname{figure}{Fig.}{Figs.}

\usepackage{subcaption}

\usepackage{placeins}
\usepackage{mathtools}

\usepackage{csquotes}
\usepackage[normalem]{ulem}

\newcommand{\etal}{\textit{et al.}} 

\usepackage{orcidlink} 

\begin{document}

\title{Resource-efficient quantum algorithms for selected  Hamiltonian subspace diagonalization}

\author{Vincent Graves ~\orcidlink{0000-0003-4856-0229}}
\email{vincent.graves@stfc.ac.uk}
\affiliation{
 National Quantum Computing Centre, RAL, Oxfordshire, United Kingdom
}
\author{Manqoba Q. Hlatshwayo ~\orcidlink{0009-0000-7977-1550}}
\affiliation{
 National Quantum Computing Centre, RAL, Oxfordshire, United Kingdom
}

\author{Theodoros Kapourniotis~\orcidlink{0000-0002-6885-5916}}

\affiliation{
 National Quantum Computing Centre, RAL, Oxfordshire, United Kingdom
}

\author{Konstantinos Georgopoulos ~\orcidlink{0000-0002-1481-2347}}
\affiliation{
 National Quantum Computing Centre, RAL, Oxfordshire, United Kingdom
}

\date{\today}

\begin{abstract}
Quantum algorithms for selecting a subspace of Hamiltonians to diagonalize have emerged as a promising alternative to variational algorithms in the NISQ era. So far, such algorithms, which include the quantum selected configuration interaction (QSCI) and sample-based quantum diagonalization (SQD), have been formulated within second-quantization in Fock space, which leads to inefficient usage of qubit resources. We introduce the first QSCI algorithm developed in the CI-matrix (CIM) framework, which is known to have optimal qubit scaling of exactly $\lceil \log_2 (N) \rceil$ where $N$ is the size of the CIM. In addition, we introduce a novel single-bit flip error mitigation which comes at the overhead of a single qubit and we combine this with a stochastic approximate Trotterization evolution adapted from qDRIFT. Simulating benchmark N$_2$ and naphthalene molecules on quantum hardware, our results achieved similar accuracy as SQD methods but with significantly less quantum resources. However, our CIM-QSCI algorithm and SQD methods could not match the performance of classical heat-bath CI (HCI) for the same task. Hence, we introduce an augmented version of QSCI called quantum selected heat-bath CI (QSHCI). This variant replaces classical heat-bath sampling with quantum sampling from QSCI to achieve performance comparable to HCI. We note that a current drawback of our approach is the preprocessing cost of $\mathcal{O}(N^2\log N)$ for constructing the CIM and performing the Pauli decomposition. This can be further improved by considering efficient CIM access models for the stochastic Trotter evolution.

\end{abstract}

\maketitle

\section{Introduction}\label{sec:intro}

Quantum chemistry has been identified as one of the high-impact use cases for quantum computing \cite{cao_quantum_2019, Bauer2020QuantumAlgorithmsQuantum, Dalzell2025QuantumAlgorithmsSurvey}. In particular, the classical computation of molecular energies via exact diagonalization becomes increasingly inefficient or intractable as the size of the molecule increases. This is because the inclusion of more electrons and orbitals in a model leads to an exponential increase in the Hilbert space. To circumvent this classical computational bottleneck, several quantum algorithms have been proposed for molecular simulation \cite{Bauer2020QuantumAlgorithmsQuantum, Dalzell2025QuantumAlgorithmsSurvey}. The quantum phase estimation (QPE) \cite{kitaev_quantum_1995} is one of the early quantum algorithms that theoretically can achieve an exponential speed-up over classical methods \cite{cao_quantum_2019}, albeit it requires large quantum resources beyond the capabilities of current quantum hardware. Hence, a set of quantum algorithms has been developed that can be run on available noisy intermediate-scale quantum (NISQ) devices. A more widely explored algorithm is the variational quantum eigensolver (VQE) \cite{Peruzzo2014VariationalEigenvalueSolver, Kandala2017HardwareEfficientVariational}, and its variants such as ADAPT-VQE \cite{Grimsley2019AdaptiveVariationalAlgorithm}. VQE-based algorithms have achieved good accuracy for small molecular problems \cite{Bauer2020QuantumAlgorithmsQuantum} but has the following limitations: (i) the parametrized circuit ansatz is susceptible to barren plateaus \cite{Larocca2025BarrenPlateausVariational}, (ii) the energy variation is not bounded below by the true ground-state energy \cite{Cerezo2021VariationalQuantumAlgorithms}, and (iii) the quantum noise can increase the computational resources due to employing error-mitigation techniques \cite{Takagi2022FundamentalLimitsQuantum, Tilly2022VariationalQuantumEigensolver}.

There are promising stochastic alternative quantum algorithms to VQE-based methods. The first is the quantum computing-enhanced quantum Monte-Carlo (QMC) method, which uses a statistical sampling of the ground state to speed up QMC \cite{Yang2021AcceleratedQuantumMonte, Huggins2022UnbiasingFermionicQuantum}. This method overcomes the sign problem faced by classical QMC for fermionic systems whilst being more robust to quantum noise than VQE, as it does not require an accurate preparation and measurement of the ground-state wavefunction. However, its success is dependent on the overlap between the initial and final states. As the overlap decreases, the number of measurements required to maintain accuracy increases exponentially \cite{Huggins2022UnbiasingFermionicQuantum}. The second is a family of sample-based Hamiltonian reduction algorithms such as the Quantum Selected Configuration Interaction (QSCI) proposed by Kanno \etal \cite{kanno_quantum-selected_2023}. This algorithm aims to stochastically reduce the size of the Hamiltonian by using the quantum computer to inform the user as to which electronic configurations contain the support of the target eigenstates, typically the ground state. The algorithm works as follows: first, an initial state is prepared on the quantum computer, which is assumed to have some non-zero overlap with the eigenstate of interest. Next, this initial state is evolved in a way that brings it closer to the true eigenstate. Subsequently, the final state is measured several times. These measurements indicate which computational basis states, or electronic configurations, are needed to recover the eigenstate of interest. The indicated configurations are then used to construct a subspace Hamiltonian, which is diagonalized classically. 

Since its advent, there have been several variants of QSCI, such as: (i) variants which uses Hamiltonian simulation instead of VQE for the approximate evolution step \cite{sugisaki_hamiltonian_2024, mikkelsen_quantum-selected_2025}; (ii) the Sample-based Quantum Diagonalization (SQD), which improved on the efficiency of the construction of the sub-space Hamiltonian through better approximate evolution and configuration recovery \cite{robledo-moreno_chemistry_2024}; (iii) extensions to compute the ground and excited states of a molecule \cite{kanno_quantum-selected_2023,barison_quantum-centric_2024,robledo-moreno_chemistry_2024}; (iv) extensions to simulation of molecular interactions \cite{kaliakin_accurate_2024}; (v) the inclusion of neural networks \cite{chen_neural_2025}; (vi) an adaptive scheme which incorporates ideas from ADAPT-VQE into QSCI resulting in ADAPT-QSCI \cite{nakagawa_adapt-qsci_2024} and many others \cite{piccinelli_quantum_2025, Shajan2025QuantumCentricSimulations, Danilov2025EnhancingAccuracyEfficiency, Kaliakin2025ImplicitSolventSample}. In addition, there have been algorithms that were inspired by QSCI, including a sample-based Krylov quantum diagonalization (SKQD) algorithm \cite{yu_quantum-centric_2025}, which combines SQD with Krylov subspace diagonalization. Throughout this paper, we refer to variants of QSCI approaches as {\it Alg}-QSCI, where {\it Alg} indicates the distinct approach used. For example, the SQD work by Robledo-Moreno \etal \cite{robledo-moreno_chemistry_2024} utilized an LUCJ ansatz; hence, we refer to it as LUCJ-SQD. 

In this work, we introduce the first class of QSCI algorithms that utilise the CI-Matrix (CIM) representation of the molecular problem in first quantisation, which we name CIM-QSCI. This representation has two major benefits, the first is an optimal qubit encoding where the number of qubits scales as $\mathcal{O}(\log_2 N)$, where $N$ is the size of the CI matrix. In contrast, previous QSCI algorithms have been formulated in second quantization, which utilizes Fock basis states, such that the number of qubits required is the number of orbitals. Second, this formalism is a common computational workflow of classical methods for quantum chemistry, hence it could efficiently leverage the synergy of a hybrid quantum-classical approach. In addition, the CIM matrix can be constructed in single-precision, which significantly saves on memory whilst achieving the same accuracy as classical double-precision algorithms. The CIM representation has been considered for quantum molecular simulations \cite{toloui_quantum_2013,babbush_exponentially_2017}, however, this is the first time it has been applied to QSCI methods.

Furthermore, the CIM-QSCI introduced here achieves a similar or better accuracy than previous QSCI algorithms whilst utilizing significantly lower quantum resources. However, this comes at a preprocessing cost of constructing the CIM before building the qubit Hamiltonian, which is an additional step second-quantised based QSCI algorithms do not require. The first-quantised CIM is then decomposed into a weighted sum of Pauli strings using the fast Walsh-Hadamard transform (FWHT) \cite{georges_pauli_2025}. Combined, these preprocessing steps add at most $\mathcal{O}(N^2\log N)$ cost to the CIM-QSCI pipeline. However, this cost can be reduced in future works by considering alternative efficient Hamiltonian access models \cite{Dalzell2025QuantumAlgorithmsSurvey}. To improve the evolution step, we utilise a stochastic approximate Trotterization scheme based on qDRIFT \cite{campbell_random_2019}, which is similar, but developed independently, to the SqDRIFT by Piccinelli \etal \cite{piccinelli_quantum_2025}. The major advantage of the approximate Trotterization is that it significantly reduces the circuit depth, making the algorithm efficient to run on currently available NISQ devices. In addition, we developed a novel single-bit-flip error mitigation scheme for the CIM framework, which provides a similar post-selection advantage as electron number conserving encodings used in second-quantised Fock state representations. We show that our error-mitigation scheme produces comparable results to the configuration recovery in second-quantisation \cite{robledo-moreno_chemistry_2024}.

However, the performance of the CIM-QSCI algorithm and the SQD approach did not reach that of the classical heat-bath configuration interaction (HCI) \cite{holmes_heat-bath_2016} method for the same problem. To address this limitation, we propose an enhanced variant of QSCI, termed quantum selected heat-bath configuration interaction (QSHCI). In this approach, the classical heat-bath sampling procedure is replaced by quantum sampling within the QSCI framework, enabling performance comparable to that of HCI. In HCI, the subspace is increased iteratively by including electronic configurations that connect to configurations already within the subspace using a deterministic heat-bath sampling \cite{holmes_heat-bath_2016}. This approach has shown to be successful at producing highly accurate eigenvalues for a low computational cost \cite{holmes_heat-bath_2016}. In CIM-QSHCI, we use a quantum-informed heat-bath sampling approach that utilises a more complex probability distribution of configurations sampled on the quantum computer.

We demonstrate our CIM-QS(H)CI algorithms on performing ground-state simulation of the molecules N$_2$ and naphthalene to benchmark against other QSCI-based algorithms \cite{kanno_quantum-selected_2023,robledo-moreno_chemistry_2024, piccinelli_quantum_2025}. The N$_2$ bond stretch problem is often used as a benchmark due to the highly-correlated nature of the problem, which makes full CI simulation necessary \cite{fan_usefulness_2006,bulik_can_2015}. In addition, the polycyclic aromatic hydrocarbon (PAH) naphthalene is studied. PAH molecules have been identified as a group of highly toxic but very stable molecules that are produced by natural sources and human activities \cite{patel_polycyclic_2020,montano_polycyclic_2025}. However, they are often difficult to describe using classical algorithms due to their aromatic nature \cite{wahab_compas_2022}. Understanding how to efficiently simulate these molecules on a quantum computer is a stepping stone to being able to simulate more complex molecules with industrial applications such as drug discovery and material science. The results show that our CIM-QSCI achieves, with significantly less quantum resources, the same accuracy as SqDRIFT \cite{piccinelli_quantum_2025} for naphthalene and better accuracy than LUCJ-SQD \cite{robledo-moreno_chemistry_2024} for N$_2$. In addition, CIM-QSHCI outperforms all types of QSCI algorithms and achieves a similar accuracy comparable accuracy and performance to HCI.

The remainder of the manuscript is structured as follows: Section~\ref{sec:method} introduces the CIM framework and associated QSCI algorithms. The computational details of the molecular simulations are presented in Section~\ref{sec:comp_details}. Finally, results from emulator and quantum hardware calculations are presented in Section~\ref{sec:results}, followed by a summary and outlook in Section~\ref{sec:conclusion}.
\section{Methodology}\label{sec:method}
\begin{figure*}[ht]
    \centering
    \includegraphics[width=\textwidth]{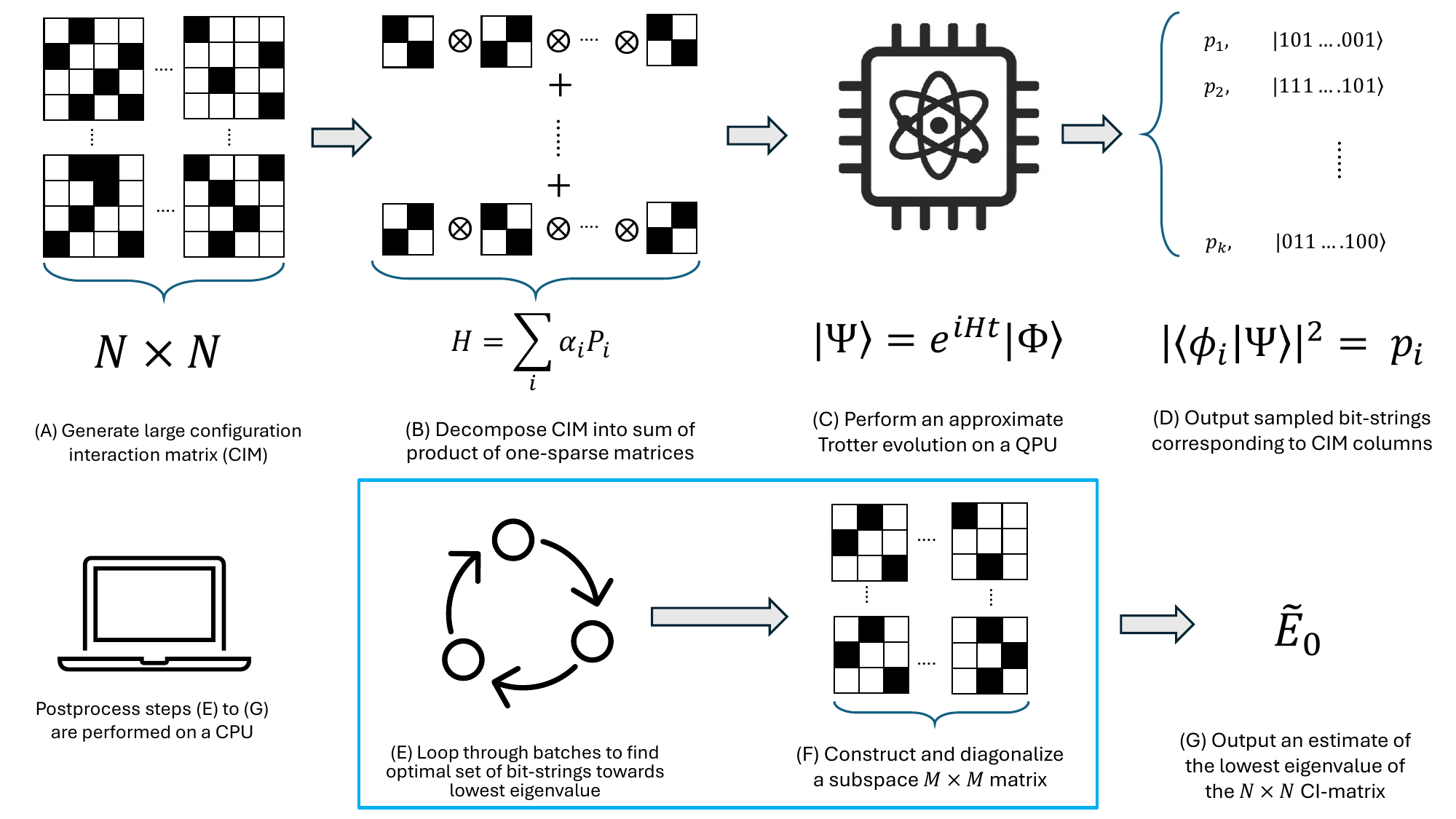}
    \caption{Overview of the QSCI algorithms based on the CIM framework for diagonalising a large general $N \times N$ matrix. For molecular chemistry problems, this matrix corresponds to the full CI-matrix.}
    \label{fig:general_CIM_QSCI}
\end{figure*}

An overview of the QSCI algorithm based on CIM framework is presented in Figure~\ref{fig:general_CIM_QSCI} and described below. This is followed by a more detailed description in the rest of this section.
\begin{enumerate}
    \item[A] \textbf{The input matrix.} The first step is to determine the matrix that has the eigenvalues of interest. In principle, any arbitrary matrix can be used as shown in Section~\ref{subsec:input-matrix}. However, in this work, the configuration-interaction matrix (CIM) is generated and used to represent the molecular eigenvalue problem. In addition, the precision of this matrix can be lower than the required accuracy of the result as this matrix isn't directly diagonalized.
    \item[B] \textbf{Decompose into one-sparse matrices.} The input matrix is then decomposed into a sum of products of Pauli matrices. This forms a qubit Hamiltonian that drives the Trotter evolution in the next step. Other decompositions and Hamiltonian access models can be used as detailed in Section~\ref{subsub:decomp-others}.
    \item[C-D] \textbf{Perform an approximate evolution} from an initial trial state to the eigenstate of interest. Here, we evolve using a modified qDRIFT Trotterization, where the circuit depth is reduced by probabilistically truncating the Lie-Trotter using the strength of the Hamiltonian terms \cite{campbell_random_2019}.
    \item[E-G] \textbf{Classical post-processing} of the sampled evolved states, outputted from a quantum computer, to construct and diagonalise a subspace-Hamiltonian to get the target eigenvalue. It is at this point that the precision of the subspace Hamiltonian needs to be appropriately high (often double-precision is sufficient). In this work, we use two approaches; first, the traditional QSCI approach \cite{kanno_quantum-selected_2023} of looping over several sampled batches to obtain the lowest subspace-eigenvalues. Second, a novel extension inspired by the heat-bath CI (HCI), which we call the Quantum Selected HCI (QSHCI).
\end{enumerate}

\subsection{Input Matrix}\label{subsec:input-matrix}

\subsubsection{A General Matrix}

In principle, any square, symmetric matrix with real values can be used as the input matrix. In addition, it can be sparse or dense, but the accuracy of the final eigenvalue(s) depends on the matrix's density. 

If the matrix is sparse, then several terms in the eigenvectors will be zero. Seeing as the aim of the QSCI-based algorithms is to solve the eigenvalue problem within a subspace, it becomes trivial to conceptualise the vectors that can be removed from the Hilbert space - those which correspond to the smallest terms in the eigenvector of interest. The problem is that one does not know which vectors will correspond to these small terms until the eigenvalue problem has been solved. 

If the matrix is fully dense, then there will be no terms in the eigenvector that are small, and as a result, using a subspace of any size will be inaccurate. Hence, QSCI-based algorithms will perform best with sparse matrices.

\subsubsection{CI-Hamiltonian Matrix}

In the results section of this work, the CI-Hamiltonian matrix (or CIM) is the matrix of interest and is, therefore, the matrix that is implemented directly on the quantum computer. This means that the approach is agnostic to the type of quantization used to construct the Slater determinants (see below), i.e, first- or second-quantization. The benefits of this representation are that it is more qubit-efficient than other quantized approaches and that using the CIM, which is sparse, improves the efficiency of state evolutions using Trotterization \cite{cao_quantum_2019}.

The CI-Hamiltonian matrix can be written in the basis of Slater determinants, which are built to be anti-symmetric with respect to the exchange of any two electrons \cite{toloui_quantum_2013}. A Slater determinant is defined as a configuration of $N_e$ electrons occupying $N_o$ orbitals. The electronic Hamiltonian matrix elements can then be computed using Slater-Condon rules \cite{babbush_exponentially_2017}, resulting in a CI-Hamiltonian matrix. Its elements can be written as,
\begin{align}\label{eq:CI_Ham}
    H_{ij} =  \left< \phi_i \right| \hat{H}^{CI} \left| \phi_j \right>
\end{align}
where $H_{ij}$ are the Hamiltonian matrix elements of $\hat{H}^{CI}$ and $\left| \phi_i \right>$ is a configuration. 

There are 4 classes of matrix elements within the Slater-Condon rules. Each class depends on the set of spin-orbitals in the configurations $\left| \phi_i \right>$: $\{\chi_k^i\}$. These classes can be written as:
\begin{itemize}
    \item[1.] If $i = j$:
    \begin{align}
        H_{ii} = \sum_{k=1}^{N_e} h_{\chi_k^i \chi_k^i} + \sum^{N_e}_{k<l} \left( h_{\chi_k^i\chi_l^i\chi_k^i\chi_l^i} - h_{\chi_k^i\chi_l^i\chi_l^i\chi_k^i} \right)
    \end{align}
    \item[2.] If $i$ and $j$ differ by one spin-orbital labelled $\chi_p^i$ and $\chi_q^j$:
    \begin{align}
        H_{ij} = h_{\chi_p^i \chi_q^j} + \sum^{N_e}_{l} \left( h_{\chi_p^i\chi_l^i\chi_q^j\chi_l^j} - h_{\chi_p^i\chi_l^i\chi_l^j\chi_q^j} \right)
    \end{align}
    \item[3.] If $i$ and $j$ differ by two spin-orbitals labelled $\chi_p^i\chi_r^i$ and $\chi_q^j\chi_s^j$:
    \begin{align}
        H_{ij} = h_{\chi_p^i\chi_r^i\chi_q^j\chi_s^j} - h_{\chi_p^i\chi_r^i\chi_s^j\chi_q^j}
    \end{align}
    \item[4.] If $i$ and $j$ differ by more than two-spin orbitals:
    \begin{align}
        H_{ij} = 0
    \end{align}
\end{itemize}
Here $h_{pq}$ and $h_{qprs}$ are one- and two-electron integrals, respectively.

An active-space is created when Slater determinants are built using a specified set of orbitals, $y$, and active electrons, $x$, instead of using the full Hilbert space. Configuration state functions (CSF) can be constructed from a linear combination of Slater determinants. Therefore, a CI-Hamiltonian can be constructed within a complete active-space (CAS), denoted ($x$, $y$). A `{\it{complete}}' active-space indicates that all excitations within the active space are included in the calculation. If all of the orbitals of a basis set are included in the CAS, then the basis set limit is reached, and the model is a full-CI (FCI).

If the symmetries of the eigenstates of interest are known, then the number of CSFs included in the CI-Hamiltonian can be reduced using the point group of the molecule. For example, N$_2$ can be described using the D$_{2h}$ point group with the ground state having a singlet A$_g$ spin-symmetry. Therefore, the CI-Hamiltonian describing the ground state only needs to include singlet A$_g$ CSFs, and CSFs of all other spin-symmetries can be discarded. 

The molecular wavefunction can be written as:
\begin{align} \label{eq:mol_wfn}
    \left| \Psi \right> = \sum_i c_i \left| \phi_i \right>,
\end{align}
where, $\left| \phi_i \right>$ is the $i$'th CSF and $c_i$ is its coefficient. As a result, the probability amplitude of state $\left| \phi_i \right>$ is equal to $|c_i|^2$. 

\subsection{Decompose Matrix into One-sparse Matrices}

In order to implement this matrix on a quantum computer, it needs to be transformed into an appropriate format. In this work, we decompose it into a set of Pauli matrices. Here, we outline the fast Walsh-Hadamard transform (FWHF) to do this, which has the best complexity scaling to date \cite{georges_pauli_2025}. Other approaches could be used and are mentioned in the following subsection.

\subsubsection{Fast Walsh-Hadamard Transform}

The CI-Hamiltonian can be decomposed into a Pauli string using the fast Walsh-Hadamard transform (FWHT) \cite{georges_pauli_2025}. This algorithm works in two stages. First, the coefficients of the Pauli strings can be computed from the Hamiltonian as:
\begin{align}
    \alpha_{r,s} = \frac{i^{-| r \land s|}}{2^q} \sum_{n=0}^{2^q - 1} h_{n \oplus r ,n} \left(H^{\otimes q} \right)_{n,s} \label{eq:fwh_coeffmat}
\end{align}
where, $\alpha_{r,s}$ is a matrix element of the coefficient matrix, $q$ is the number of qubits, $H$ is the Hadamard matrix and $h_{i,j}$ are the CI-Hamiltonian matrix elements with index $i,j$. The bitwise operators $\land$ and $\oplus$ indicate AND and XOR operations. This coefficient matrix requires the full CI-Hamiltonian to be known. The Pauli string can then be computed as:
\begin{align}
    P_{r,s} = \bigotimes_{j=0}^{q-1} i^{r_j \land s_j} X^{r_j} Z^{s_j} 
    \label{eq:fwh_pauli}
\end{align}
Where, here $I, X, Y$ and $Z$ are Pauli matrices. Combining these two steps results in the qubit-Hamiltonian:
\begin{align}
    \hat{H}_q = \sum_{r,s} \alpha_{r,s} P_{r,s} \label{eq:fwh_all}
\end{align}
This approach scales as $O(2^{2q} \log{(2^q)})$ where $q$ is the number of qubits. The number of qubits required for this encoding is dependent on the size of the Hamiltonian and is given as: $q = \lceil \log{_2(N_{CSF})} \rceil$ where $N_{CSF}$ is the number of CSF (the size of the Hamiltonian). Therefore, in terms of the number of CSF of the model, the FWHT scales as $O\left(N_{CSF}^2\log N_{CSF} \right)$.

This decomposition technique could impact the potential for quantum advantage due to its scaling. In fact, the decomposition of the CI-Hamiltonian into a Pauli string is limited by a scaling of $O\left(N_{CSF}^2\right)$ as each matrix element needs to be evaluated and there are $N_{CSF}^2$ matrix elements. As a result, the use of the FWHT adds the cost of $O\left(\log N_{CSF}\right)$ when compared to a (theoretically) optimal technique. In addition, other decomposition techniques have been proposed which have similar scaling \cite{hamaguchi_handbook_2024, hantzko_tensorized_2024}. Of note is the Tensorized Pauli Decomposition (TPD) technique of Hantzko \etal \cite{hantzko_tensorized_2024} which also scales as $O\left(N_{CSF}^2\log N_{CSF} \right)$. There are computational implementations of both the FWHT and TPD, however, it has been shown that the FWHT implementation is more efficient and as a result, quicker \cite{georges_pauli_2025}. Hence, we use the FWHT in this work. We emphasise that there is a trade-off between the scaling of this step and the reduction in the number of qubits versus when traditional second-quantized approaches are used. Furthermore, in second-quantized approaches, the qubit-Hamiltonian still has to be generated. The only additional step in this work is the generation of the CI-Hamiltonian.

\subsection{Approximate Trotter Evolution}\label{sec:qubitHam_evol}

Here, we use an approximate Trotterization approach to evolve the initial trial wavefunction. Our trial wavefunction is simply the Hartree-Fock (HF) determinant. A benefit of this is that there is significantly less (classical) pre-processing required than the LUCJ-SQD approach, which necessitates a CCSD calculation as the pre-processing step \cite{robledo-moreno_chemistry_2024}. CCSD calculations scale as $N_e^6$ \cite{mahajan_beyond_2025} (where $N_e$ is the number of electrons) and are often sufficiently accurate, removing the need for the quantum algorithm to begin with. 
Using the HF state as the initial state assumes that it has a sufficiently good overlap with the true ground state and, therefore, that the evolution will result in the desired state. A similar assumption has been used in previous work for weakly-correlated systems \cite{tubman_postponing_2018}. The preparation of an initial state that has a sufficient overlap for highly correlated systems is a non-trivial, open problem and can be the most expensive part of an algorithm \cite{tubman_postponing_2018, scali_purified_2025}.

Sugisaki \etal \cite{sugisaki_hamiltonian_2024} proposed a Hamiltonian simulation-based QSCI approach (HSB-QSCI) which utilised a (full) Trotterization evolution. It was shown that, for the systems they studied, a short-time evolution of $t = 1$ was sufficient to collect the important configurations. However, the use of Lie- or Suzuki-Trotter Trotterization can result in very deep circuits, even for short evolution times. As such, these circuits are often far too long to be run on near-term quantum devices. To overcome this, Piccinelli \etal \cite{piccinelli_quantum_2025} proposed a QSCI-type approach which utilised an approximate Trotterization based on qDRIFT by Campbell \cite{campbell_random_2019}. This approach, called SqDRIFT, actually performs a variation of Sample-based Krylov Quantum Diagonalization (SKQD) \cite{yu_quantum-centric_2025} instead of QSCI, which is a similar sub-space diagonalization approach.

In this work, we propose an approximate evolution utilising qDRIFT, similar to SqDRIFT and to the work of Sugisaki \cite{sugisaki_size-consistent_2025}, but developed independently, around the same time.

\subsubsection{qDRIFT and its Modifications}

In qDRIFT Trotterization, the circuit depth is reduced by probabilistically truncating the Lie-Trotter Trotterization using the strength of the Hamiltonian terms. i.e, for a Hamiltonian written as a Pauli string, as in Equation~\ref{eq:fwh_all}, the Hamiltonian that is Trotterized is a truncated version of $\hat{H}_q$ where the terms that have the smallest $\alpha_{r,s}$ are probabilistically removed. This is done using a weighted probabilistic formalism based on the magnitude of $\alpha_{r,s}$. Details of this approach can be found in \cite{campbell_random_2019}. Note that because the CI-Hamiltonian is symmetric and real, all $\alpha_{r,s}$ values are real.

To retain a reasonable accuracy, Campbell \cite{campbell_random_2019} proposed that the number of terms sampled in the time evolution, $n_c$, should be:
\begin{align}\label{eq:qdrift_terms}
    n_{c} = \frac{2 \lambda^2 t^2}{\epsilon}
\end{align}
where $\lambda = \sum_{r,s} \left|\alpha_{r,s} / \max{(\alpha)}\right|$, $t$ is the time evolution used in the Trotterization, $\epsilon$ is the desired accuracy and $\max(\alpha)$ is the largest $|\alpha_{r,s}|$ value for all $r$ and all $s$. Note that each term is chosen independently and as a result, duplicate choices are possible. 

In this work, we propose an Approximate Evolution (AE) technique, which further truncates the Trotterization. This is done using $n_a$ instead of $n_c$, where,
\begin{align}
    n_{a} = 2\lambda t^2 \label{eq:qdrift_ourN}
\end{align}
This significantly reduces the computational requirements but comes at the cost that the Hamiltonian evolution is less accurate. To overcome this, the evolution is run several times, and the bitstrings from each run are combined into a single set. As a result, it follows that there should be $r$ repetitions such that,
\begin{align}
    r n_{a} & =  2\lambda^2 t^2 \label{eq:qdrift_ourR}
\end{align}
This ensures that the whole Hilbert space can still be explored while keeping the computational costs down. We implement a single time step ($t = 1$), however, we note that for some systems, i.e., those with several low-lying states that are very close in energy, $t = 1$ may not be sufficient to ensure the important configurations are collected. One approach could be to perform the QS(H)CI calculation for several time durations until the result converges. In this work, we determined that a single time step is sufficient; however, we cannot rule out the need for $t > 1$ for other molecules or problems. In practice, an integer of $r$ and $n_a$ are needed and hence they are both rounded up.

Conceptionally, Equation~\ref{eq:qdrift_terms} and Equation~\ref{eq:qdrift_ourR} are equivalent when $r = \epsilon$. This equivalence is discussed, within the context of qDRIFT combined with SKQD, in the work of Piccinelli \etal \cite{piccinelli_quantum_2025}. Here, the algorithm presented is a special case of SKQD equivalent to when $d$, the Kylov subspace order, is equal to 2. As a result, the proof of the error bound in \cite{piccinelli_quantum_2025}, which shows that the approximate Trotterization based on qDRIFT is sufficient to recover the ground state, holds. For conciseness, we direct the reader to Lemma A.1 of \cite{piccinelli_quantum_2025} for a detailed proof.

We argue that this is sufficient for the QSCI-based approaches because we assume that the eigenvalue of interest is concentrated, i.e, it only has a small number of support vectors relative to the whole vector space. In addition, the sampled bitstrings only need to represent the approximate molecular wavefunction (Equation~\ref{eq:prob_dist}). This is because the post-processing step will be performed several times with the aim of recovering the accuracy required. As a result, a less accurate Hamiltonian evolution is still sufficient to recover the {\it important} bitstrings. This is similar to what was seen by Sugisaki \etal who found that $t = 1$ was sufficiently accurate \cite{sugisaki_hamiltonian_2024} and by Piccinelli \etal \cite{piccinelli_quantum_2025} who also used a version of qDRIFT. If one wanted to accurately simulate the Hamiltonian evolution, for example, in QPE, then this AE would result in a large error and would be inappropriate. 

We also note that the scaling of the $\lambda$ parameter in Equation~\ref{eq:qdrift_ourN} and Equation~\ref{eq:qdrift_ourR} could make this approach intractable for larger problem sizes. We propose that this could be overcome via the qSWIFT \cite{nakaji_high-order_2024} and qSHIFT \cite{lee_qshift_2026} variations of qDRIFT which remove the dependence on $\lambda$. However, we leave this to future work. 

\subsubsection{Alternative Approach}\label{subsub:decomp-others}

An alternative to the FWHT and Trotterization is to use the graph colouring algorithm of Babbush \etal \cite{babbush_exponentially_2017}. In this algorithm, the CIM is decomposed into a sum of one-sparse Hamiltonians, each of which contains a single molecular integral. This is further decomposed by discretizing the integrals in real space into a sum of self-inverse operators to produce the circuit \texttt{SELECT($\hat{H}$)}. By constructing a quantum circuit that repeatedly calls \texttt{SELECT($\hat{H}$)}, the initial quantum state is evolved in a similar way as Trotterization. The total time complexity of this approach is $\tilde{\mathcal{O}}(N_e^2N_o^3t)$ where $N_e$ is the number of electrons, $N_o$ is the number of orbitals, and $t$ is the evolution time. The number of qubits is $\tilde{\mathcal{O}}(N_e)$. Note the $\tilde{\mathcal{O}}$ indicates an asymptotic upper bound suppressing any polylogarithmic factors. In this work, we utilise the FWHT with Trotterization and leave the graph colouring algorithm for future work.

\subsection{Post-selection Error Mitigation}

\subsubsection{Bit-flip Mitigation} \label{sec:bit-flip}

The output of the quantum computer is a series of sampled bitstrings, each one corresponding to a specific CSF. These are used to construct a subspace-Hamiltonian (see Section~\ref{sec:subspaceHam}). On a noisy device, some of the sampled bitstrings may have undergone bit-flip errors, which may broaden the distribution of sampled bitstrings and result in undesired CSFs in the subspace-Hamiltonian. 

To overcome this, we impose a bit-flip mitigation scheme that works by ensuring that the bits in the binary representations of the CSFs always sum to odd/even numbers. This means that a single bit-flip from a physical state will result in a non-physical state, which can be easily post-selected out. As a result, we ensure that the Hamming distance between bitstrings of physical states is always a multiple of 2. An example of how this works is as follows. Suppose we have two qubits and, therefore, four CSFs labelled: $\left|\phi_0\right>$, $\left|\phi_1\right>$, $\left|\phi_2\right>$, $\left|\phi_3\right>$. The corresponding bitstring representations, without the bit-flip mitigation scheme, would be:
\begin{itemize}
    \item $\left|\phi_0\right>$ : $\left|00 \right>$
    \item $\left|\phi_1\right>$ : $\left|10 \right>$
    \item $\left|\phi_2\right>$ : $\left|01 \right>$
    \item $\left|\phi_3\right>$ : $\left|11 \right>$
\end{itemize}
Now, suppose we performed a measurement on the quantum computer and the measured bitstring was meant to be $\left|00 \right>$, but a bit-flip error resulted in the measurement of $\left|01 \right>$. This would mean that the bit-flip error resulted in the incorrect identification of the physical state $\left|\phi_2\right>$ instead of $\left|\phi_0\right>$. This could result in $\left|\phi_2\right>$ being wrongly included in the subspace Hamiltonian, wasting computational expense.

Now, suppose we implement the bit-flip mitigation scheme. Here, the bitstring representations of the CSFs would be:
\begin{itemize}
    \item $\left|\phi_0\right>$ : $\left|000 \right>$
    \item $\left|\phi_1\right>$ : $\left|011 \right>$
    \item $\left|\phi_2\right>$ : $\left|101 \right>$
    \item $\left|\phi_3\right>$ : $\left|110 \right>$
\end{itemize}
If the measured bitstring was meant to be $\left|000 \right>$ but underwent a bit-flip error, then $\left|001 \right>$ might be measured instead. Seeing as $\left|001 \right>$ is a non-physical state, and therefore does not correspond to a CSF, we can know that a bit-flip error has occurred and discard this measurement from the results.

In practice, this increases the qubit requirement from $\lceil \log{_2(N_{CSF})} \rceil$ by 1 but allows for a post-selection step which removes bitstrings that are known to have bit-flip errors. This results in a similar level of bit-flip mitigation as proposed in other works, which use Fock states \cite{kanno_quantum-selected_2023, robledo-moreno_chemistry_2024}. In that case, the sum of the bitstrings should equal the number of electrons in the system. If a bit-flip error occurs, then the sum would not be as expected, and so the measurement would be discarded.

Utilising this bit-flip mitigation scheme results in several of the measured bitstrings being flagged as containing a bit-flip error and as a result, removed from the sampled space. This means that the number of retained samples will not be equal to the number of shots performed, leading to wasted computational power. To overcome this, we extend the bit-flip mitigation scheme to include the recovery of non-physical bitstrings.

Using the bit-flip mitigation scheme we could, as an example, have the same physical-states as in the example above ($\left|\phi_0\right>$, $\left|\phi_1\right>$, $\left|\phi_2\right>$, $\left|\phi_3\right>$).
Let's suppose the non-physical state $\left|010 \right>$ was measured from one shot. This non-physical state has a Hamming distance of 1 from the physical states $\left|\phi_0\right>$, $\left|\phi_1\right>$, and $\left|\phi_3\right>$ but a Hamming weight of 3 from $\left|\phi_2\right>$. As a result, assuming it was measured as a result of a single bit-flip error, it was supposed to be either $\left|\phi_0\right>$, $\left|\phi_1\right>$ or $\left|\phi_3\right>$. In this example, lets suppose that $\left|\phi_0\right>$ was measured more often than $\left|\phi_1\right>$ or $\left|\phi_3\right>$. In this case, we would add the non-physical measurement of $\left|010 \right>$ to the physical state $\left|\phi_0\right>$. 

Using this scheme, the measured non-physical states can be easily corrected and included in the sampled space. As a result, significantly more of the shot results are included in the measured space and are not thrown away.

Further analysis of this bit-flip mitigation scheme is presented in the Appendix \ref{sec:bit_flip_analysis}.

\subsection{Subspace-Hamiltonian}\label{sec:subspaceHam}

\subsubsection{Quantum Selected CI}

On the quantum computer, a probability distribution proportional to the desired wavefunction, $\left|\Psi\right>$, is measured in the computational basis as,
\begin{align}
    P_{s}(i) \propto | c_i |^2 \label{eq:prob_dist}
\end{align}
where $P_s(i)$ is the sampled probability of configuration $i$ and $c_i$ is the molecular wavefunction coefficient (see Equation ~\ref{eq:mol_wfn}).
The set of sampled CSF, $S$, then undergoes the bit-flip mitigation post-selection process described above.

The subspace-Hamiltonian, $\hat{H}^{CI}_{S}$, is then constructed as,
\begin{align}
    \hat{H}^{CI}_{S} = \hat{P}_{S} \hat{H}^{CI} \hat{P}_{S} \label{eq:subspace_ham}\\
    \hat{P}_{S} = \sum_{x\in S} \left|x\right> \left<x\right|
\end{align}
This step can be repeated several times, taking the post-selected CSFs in batches. Doing this allows more of the sampled space to be analyzed. This step could be performed in two ways: the first is to compute $\hat{H}^{CI}_{S}$ as a subset matrix of $\hat{H}^{CI}$ directly. Using this approach would result in $\hat{H}^{CI}_{S}$ inheriting the same level of precision as $\hat{H}^{CI}$. However, in our implementation, we compute $\hat{H}^{CI}$ in single precision but want $\hat{H}^{CI}_{S}$ in double precision. As a result, we (re)compute all of $\hat{H}^{CI}_{S}$ on-the-fly in double precision for each subspace.

The size of the subspace-Hamiltonian is chosen by the user. In practice, one could diagonalize the largest subspace-Hamiltonian they can on their computer, as this gives the best probability of finding accurate eigenvalues. In this work, we show the impact of choosing different subspace-Hamiltonian sizes. 

Finally, the eigenvalues of the subspace-Hamiltonian are determined exactly via diagonalization of $\hat{H}^{CI}_{S}$ (on a classical computer).

The construction and diagonalization of the subspace-Hamiltonian can be done several times, providing a range of eigenvalues. As the approach outlined here is within the variational-principle, the most accurate eigenvalues will be those that are the lowest in energy \cite{mikkelsen_quantum-selected_2025}.

\subsubsection{Quantum Selected Heat-bath CI}\label{sec:qshci_method}

The quantum selected heat-bath CI (QSHCI) algorithm follows the same workflow as heat-bath CI (HCI), but with a modified acceptance criterion. The algorithm has the following steps:
\begin{itemize}
    \item[1.] The subspace is initialized with only the initial trial wavefunction.
    \item[2.] Compute the eigenvector of the current subspace.
    \item[3.] Add configuration $k$ to the subspace if:
    \begin{align}
        \left| H_{ki}c_i \right| > \frac{\sqrt{P_k}}{v}
    \end{align}
    where configuration $i$ is already in the subspace, $H_{ki}$ is the matrix element with index $[k,i]$, $c_i$ is the coefficient of configuration $i$ in the current eigenvector, and $P_k$ is the probability that configuration $k$ was sampled on the quantum computer. The term $v$ is a user-defined variance factor which relaxes the acceptance criteria if desired. By default, this factor should be equal to 1.0.
    \item[4.] Repeat steps 2. and 3. until no additional configurations are added to the subspace.
\end{itemize}
Once the subspace has converged, the final eigenvalue is taken as the result.

In HCI, the same workflow is used except that the acceptance criteria is:
\begin{align}
    \left| H_{ki}c_i \right| > \epsilon
\end{align}
where $\epsilon$ is a user-defined tolerance.

Conceptually, QSHCI functions the same as HCI but removes the user-defined tolerance and instead introduces a variable acceptance criterion that changes for every configuration $k$. This acceptance criterion is informed by the probability distribution of configurations measured on the quantum computer.

\subsection{Complexity Analysis}

The highest computational cost of the presented algorithms comes from the FWHT, which has a scaling of $O\left(N_{CSF}^2\log N_{CSF} \right)$. This compares unfavourably for diagonalization of sparse matrices when using an iterative diagonalization technique; $O\left(N_{CSF}^2 \right)$ but better with exact diagonalization, which scales as $O\left(N_{CSF}^3 \right)$. However, the formulation presented here retains the potential for advantage against exact diagonalization in three main areas:
\begin{itemize}
    \item[1.] The number of qubits required here is exactly $\lceil \log_2 (N_{CSF})+1 \rceil$, which is less than the traditional second-quantized approach, which scales in the number of qubits as the number of spin-orbitals. In addition, the use of molecular point groups allows us to remove many determinants {\it a priori} that do not contribute to the eigenvector of interest at the point of the Hamiltonian construction. Previous works have not utilised the point group symmetry and therefore have a larger space to search over to start with.
    \item[2.] The memory requirement for the input matrix is only single-precision, which is half the standard double-precision requirements. Note that the final subspace Hamiltonian is still constructed in double-precision so that the final eigenvalues inherit this higher level of precision. An additional point is that this reduced-precision approach is not specific to our problem formulation. More traditional, second-quantized, approaches could also utilise a lower precision level when computing their qubit Hamiltonian. The precision of the Hamiltonian is only important at the point of computing the eigenvalues (the diagonalization step).
    \item[3.] The use of the approximate evolution means that only the Pauli strings that are used are computed, and those with a small coefficient are never executed on the quantum computer. In addition, this approach results in a lower gate count than full Trotterization which can easily become intractable.
\end{itemize}

The post-processing diagonalization step of the QSHCI algorithm presented here scales the same as (traditional) HCI \cite{holmes_heat-bath_2016}.

\section{Computational Details} \label{sec:comp_details}

\begin{table*}[t]
\centering
\renewcommand{\arraystretch}{1.3}
\caption{Quantum resource requirements of our CIM-based QSCI algorithms compared to SQD and SqDRIFT methods in second quantisation. The gates are the average number of 2-qubit gates, and the depth is the average 2-qubit gate depth (rounded to the nearest gate). For the molecular model, the active space comprises the number of electrons and orbitals, and the cc-pVDZ basis set was used throughout. The evolution parameters for CIM-QSCI, CIM-QSHCI, and SqDRIFT show a range of values for the ($n_a, r$), where $n_a$ is the number of Pauli terms retained in each of the $r$ repetitions of the approximate evolution. For LUCJ-SQD, the evolution parameters are the number of LUCJ layers used in the ansatz. The approximation degree (AD) is the circuit approximation by the transpiler (see Appendix \ref{sec:approx-transp}). The method parameters for CIM-QSCI, LUCJ-SQD, and SqDRIFT are the percentage size of CIM, whereas for CIM-QSHCI, it is the variance factor. Results with a superscript $\dagger$ are results from \cite{piccinelli_quantum_2025}.}

\begin{tabular*}{\textwidth}{@{\extracolsep{\fill}} l c c cc c ccc c}
\toprule
\multirow{2}{*}{Method} 
& \multirow{2}{*}{Molecule} 
& \multirow{2}{*}{Active Space}
& \multicolumn{2}{c}{Circuit Parameters} 
& Method
& \multicolumn{3}{c}{Quantum Resources}
& \multirow{2}{*}{Results} \\

\cmidrule(lr){4-5} \cmidrule(lr){7-9}

& & & Evolution & AD 
& Parameters
& Qubits & Gate & Depth & \\

\hline\hline

\multirow{5}{*}{CIM-QSCI}
& \multirow{2}{*}{N$_2$} & (10, 10) & (19 - 27, 10 - 14) & 1.00 & 40, 60, 80 & 14 & 716 & 596 & Fig~\ref{fig:fig1-CIM_QSCI_N2_Results} \\ 
& & (10, 12) & (254 - 356, 127 - 178) & 0.50 & 40, 60, 80 & 18 & 43 & 23 & Fig~\ref{fig:fig2-CIM_QSCI_N2_Results} \\
\cmidrule{2-10}

& \multirow{3}{*}{Naphthalene}
& (10, 10) & (8, 4) & 1.00 & 10 - 90 & 14 & 82 & 64 & Fig~\ref{fig:nap_ham_size} \\
& & (10, 10) & (10, 100) & 1.00 & 10 - 90 & 14 & 128 & 94 & Fig~\ref{fig:nap_ham_size} \\
& & (10, 10) & (20, 200) & 1.00 & 10 - 90 & 14 & 278 & 211 & Fig~\ref{fig:nap_ham_size} \\

\hline\hline

\multirow{4}{*}{CIM-QSHCI}
& \multirow{2}{*}{N$_2$}
& (10, 10) & (19 - 27, 10 - 14) & 1.00 & 1, 100 & 14 & 716 & 596 & Fig~\ref{fig:fig1-CIM_QSHCI_N2_Results} \\ 
& & (10, 12) & (254 - 356, 127 - 178) & 0.50 & 1 & 18 & 43 & 23 & Fig~\ref{fig:fig2-CIM_QSHCI_N2_Results} \\
\cmidrule{2-10}

& \multirow{2}{*}{Naphthalene}
& (10, 10) & (8, 4) & 1.00 & 100 & 14 & 82 & 64 & Fig~\ref{fig:nap_ham_size} \\
& & (10, 10) & (20, 200) & 1.00 & 100 & 14 & 278 & 211 & Fig~\ref{fig:nap_ham_size} \\

\hline\hline

\multirow{3}{*}{LUCJ-SQD \cite{robledo-moreno_chemistry_2024}}
& N$_2$ & (10, 10) & 2 & 1.00 & 50 & 20 & 839 & 166 & Fig~\ref{fig:fig1-CIM_QSCI_N2_Results} \\ 
& Naphthalene$^\dagger$ & (10, 10) & 1 & 1.00 & 10 - 90 & 20 & 324 & - & Fig~\ref{fig:nap_ham_size} \\
& Naphthalene & (10, 10) & 2 & 1.00 & 10 - 90 & 20 & 804 & 130 & Fig~\ref{fig:nap_ham_size} \\

\hline\hline

\multirow{2}{*}{SqDRIFT \cite{piccinelli_quantum_2025}}
& Naphthalene$^\dagger$ & (10, 10) & (10, 1000) & 1.00 & - & 20 & 113 & 69 & - \\
& Naphthalene$^\dagger$ & (10, 10) & (10, 500) & 1.00 & 10 - 90 & 20 & - & - & Fig~\ref{fig:nap_ham_size} \\

\bottomrule
\end{tabular*}
\label{tab:all_details}
\end{table*}

\subsection{Molecular parameters}

\textbf{N$_2$ Molecule}: The N$_2$ bond stretch calculations were performed for bond lengths in the range of 0.7 to 3.0 \AA. Emulator runs were performed with 24 distances with an interval of 0.1 \AA. However, hardware calculations were performed with 9 distances, which recovered the overall shape of the potential energy curve. The basis set was cc-pVDZ, and the active space was a (10,10) for the emulated results and a (10,12) for the hardware runs. This corresponded to singlet A$_g$ CSFs of about 7,992 to 8,072 and 78,832 to 78,840, respectively, for the (10,10) and (10,12) active spaces, depending on the interatomic distance. As the eigenstate of interest was of the ground state and the N$_2$ molecule has a molecular Abelian point group of D$_{2h}$, only A$_g$ CSFs were needed in the Hamiltonian. 

\textbf{Naphthalene}: The geometry of the naphthalene molecule was taken from Ghosh \etal \cite{ghosh_generalized-active-space_2017} with a point group of D$_{2h}$. The cc-pVDZ basis set was used with a (10,10) active space, replicating the model used by Piccinelli \etal \cite{piccinelli_quantum_2025}. The aim here was to compute the ground state energy, which will correspond to the lowest singlet A$_g$ state. As a result, all 7,992 singlet A$_g$ CSFs were included in the construction of the CI-Hamiltonian matrix.

For both molecules, the values of circuit parameters and quantum resources are summarized in Table~\ref{tab:all_details}.

\subsection{Implementation}

\textbf{Input matrix}: The first step of the CIM-QS(H)CI is to generate the input Hamiltonian. First, Psi4 \cite{smith_psi4_2020} was used to perform a Hartree-Fock calculation, which generated the orbitals and electron integrals needed to construct the CI-Hamiltonian matrix. The PyCI package \cite{richer_pyci_2024} was used to construct the CI-Hamiltonian with CSFs of the appropriate spin-symmetries. This CI-Hamiltonian was computed at single precision and therefore requires half as much memory as the standard, double-precision CIM, which is typically used in quantum chemistry calculations. This was possible because the full CI-Hamiltonian was not diagonalized directly and thus does not require high precision. 

\textbf{Decompose into qubit Hamiltonian}: The coefficient matrix of the qubit Hamiltonian (Equation~\ref{eq:fwh_coeffmat}) was then computed using a modified version of the {\it pauli\_lcu} library git repository \cite{georges_pauli_2025}. Similarly, these matrices were constructed in single precision. The number of terms $n_a$ of the truncated qDRIFT Hamiltonian (Equation~\ref{eq:qdrift_ourN}) was probabilistically chosen from the coefficient matrix, hence the only Pauli strings computed (Equation~\ref{eq:fwh_pauli}) were the $n_a$ strings used to construct the truncated qubit Hamiltonian (Equation~\ref{eq:fwh_all}). 

\textbf{Approximate Trotter evolution}: The Trotterized quantum circuit for this qubit Hamiltonian was then constructed as described in Section~\ref{sec:method}. This was repeated $r$ times, and the generated quantum circuits were run on an emulator or quantum hardware. All computations on an emulator were performed using the Qiskit AerSimulator with 10,000 shots and an {\it IBM FakeMarrakesh} noise model \cite{noauthor_compute_2026}. Quantum hardware calculations were performed on the Rigetti 82-superconducting-qubit device {\it Ankaa-3} via the AWS Braket service \cite{noauthor_rigetti_nodate}. Both of these QPUs have a square-array architecture.

In the case of deep quantum circuits that could not be supported by Rigetti's {\it Ankaa-3}, a circuit approximation using Qiskit transpiler was utilised \cite{cross_validating_2019}. This was tested across various instances as described in Appendix~\ref{sec:approx-transp}, and the approximation degree (AD) of 0.5 was found sufficient to obtain shallower circuits with approximately the same target evolution.

\textbf{Classical post-processing}: The sampled bitstrings were combined into a single set for post-processing. First, we postselected using the bit-flip mitigation scheme described in Section~\ref{sec:bit-flip}. For CIM-QSCI, we run the post-processing loop 10 times with 100 batches of sampled bitstrings, which yield a subspace Hamiltonian constructed in double-precision. For CIM-QSHCI, the post-processing loop was performed as described in Section~\ref{sec:qshci_method}. The sub-space Hamiltonians generated were then diagonalized using the Lanczos method provided within PyCI to get the ground-state energies reported in the results Section~\ref{sec:results}.

For benchmarking with our CIM-QS(H)CI methods, molecular simulations employing LUCJ-SQD methods were performed using the SQD-Qiskit addon \cite{noauthor_qiskitqiskit-addon-sqd_2025}. CCSD calculations performed with PySCF \cite{sun_recent_2020} were used as input to the LUCJ ansatz, which had two repetitions. 
\section{Results and Analysis} \label{sec:results}

This section presents the molecular simulation results and analyses of our CIM-based algorithms, benchmarked against other quantum algorithms reported in the literature and classical methods.  As described in Section~\ref{sec:comp_details}, the simulations were performed on an emulator with a quantum-device noise model, and the hardware calculations were on Rigetti's {\it Ankaa-3} QPU. The quantum resources requirements for our CIM-based algorithms compared to SQD and SqDRIFT methods are shown in Table~\ref{tab:all_details}. The simulation results show that our CIM-based algorithms achieve comparable or better accuracy than SQD methods while using lower quantum resources.

\subsection{CIM-QSCI for simulating N$_2$} \label{subsec:CIM-QSCI-N2}

\begin{figure*}[t]
\centering

\begin{subfigure}{0.49\textwidth}
    \centering
    \includegraphics[width=\linewidth]{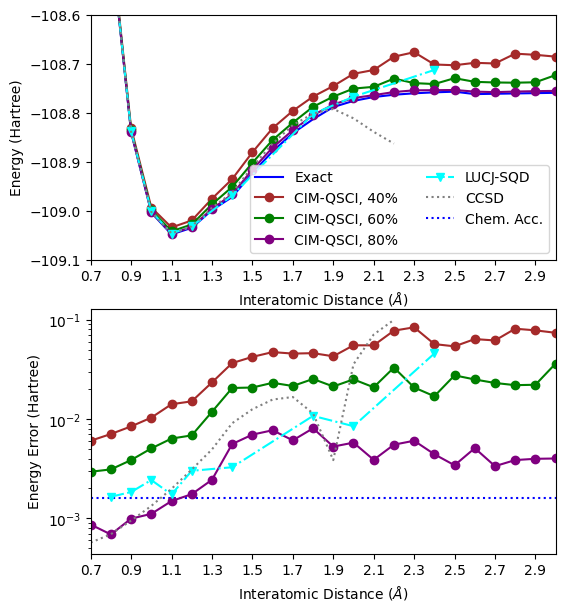}
    \caption{Emulator results for molecular simulation performed using a (10,10) active space. }
    \label{fig:fig1-CIM_QSCI_N2_Results}
\end{subfigure}
\hfill
\begin{subfigure}{0.49\textwidth}
    \centering
    \includegraphics[width=\linewidth]{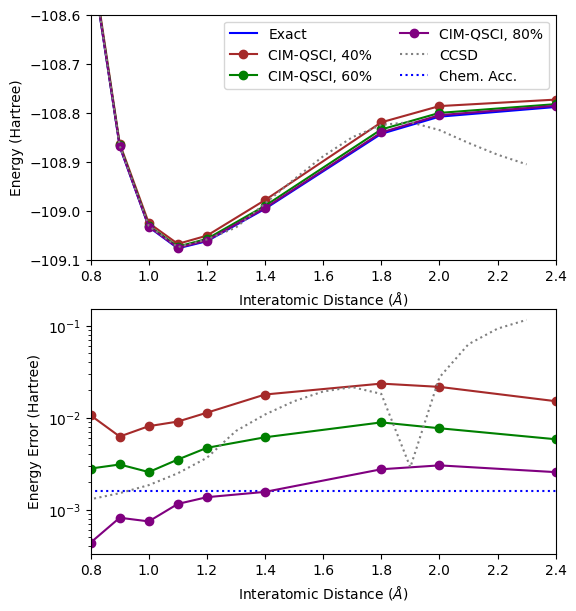}
    \caption{Quantum hardware results for molecular simulation performed using a (10,12) active space. }
    \label{fig:fig2-CIM_QSCI_N2_Results}
\end{subfigure}

\caption{Simulated potential energy curve of N$_2$ molecule (upper panel) and absolute error of the methods (lower panel) relative to exact diagonalization for different bond lengths. The results of our CIM-QSCI are shown as brown, green, and purple solid lines with circles, corresponding to 40\%, 60\%, and 80\% of the subspace Hamiltonian, respectively. These results are compared to CCSD, LUCJ-SQD with two LUCJ repetitions (on emulator), and exact diagonalization computed using a Lanczos method. All error curves (bottom panel) are plotted against chemical accuracy (0.0016 Hartree) shown as a blue dotted line. }
\label{fig:CIM_QSCI_N2_Results}
\end{figure*}

The potential energy curve for N$_2$ with various bond lengths computed using CIM-QSCI is shown in Figure~\ref{fig:CIM_QSCI_N2_Results} along with the corresponding energy errors relative to exact diagonalization.

We note that the overall shape of the potential energy curve from both the emulator and quantum hardware matches the exact curve when using the CIM-QSCI algorithms, regardless of the size of the subspace Hamiltonian. In contrast, the CCSD method deviates from the exact curve from around 1.8 \AA~and fails to converge beyond 2.3 \AA. 

We observe that the accuracy of the CIM-QSCI algorithm decreases with longer interatomic distances, as shown in the lower panels of Figure ~\ref{fig:CIM_QSCI_N2_Results}. The error in CIM-QSCI increases until it plateaus at around 1.4 \AA~and 1.8 \AA~for the emulator and hardware results, respectively. This is because, as the interatomic distance increases, the correlation energy also increases until the bond-dissociation limit. At the point that the correlation energy peaks, the overlap between the Hartree-Fock wavefunction (the initial state) and the true ground state wavefunction will be smallest. Hence, the CIM-QSCI algorithm performs worse because its success depends on this overlap. In addition, as the correlation energy increases, the number of determinants needed to recover the true ground-state wavefunction will increase, meaning that the size of the subspace Hamiltonian will need to be larger. The average error for the CIM-QSCI with 40\%, 60\%, and 80\% of the full-sized Hamiltonian on the emulator was 0.045, 0.019, and 0.0041 Hartree, respectively, whilst on the quantum hardware it was 0.014, 0.0050, and 0.0016 Hartree, respectively. Most of these average error results are slightly above chemical accuracy, which reflects the capabilities of current NISQ hardware, which we expect to improve as the technology matures.

For the results using the (10,10) active space, the CIM-QSCI algorithm, using 80\% of the full-sized Hamiltonian, performs slightly better than the LUCJ-SQD result, particularly for larger interatomic distances. This could be due to the LUCJ-SQD algorithm using excitation functions from CCSD calculations, which have been shown to fail at large interatomic distances. In addition, the size of the subspace Hamiltonian for LUCJ-SQD was set to 50\% which is smaller than the 80\% CIM-QSCI work presented here.

\subsection{CIM-QSHCI for simulating N$_2$}\label{subsec:CIM-QSHCI-N2}

\begin{figure*}[t]
\centering

\begin{subfigure}{0.49\textwidth}
    \centering
    \includegraphics[width=\linewidth]{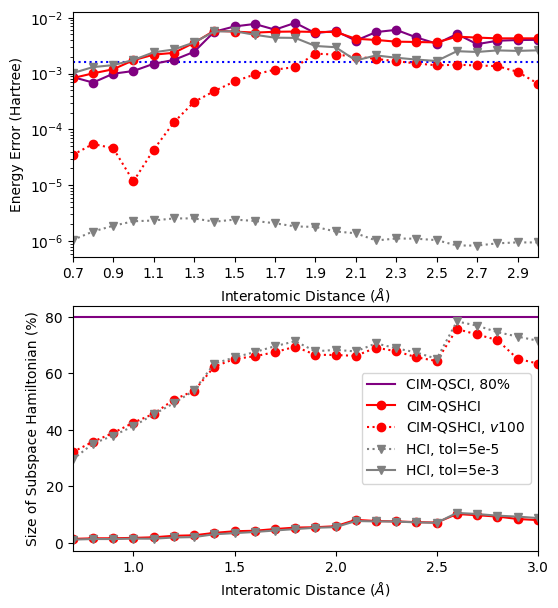}
    \caption{Emulator results for molecular simulation performed using a (10,10) active space. }
    \label{fig:fig1-CIM_QSHCI_N2_Results}
\end{subfigure}
\hfill
\begin{subfigure}{0.49\textwidth}
    \centering
    \includegraphics[width=\linewidth]{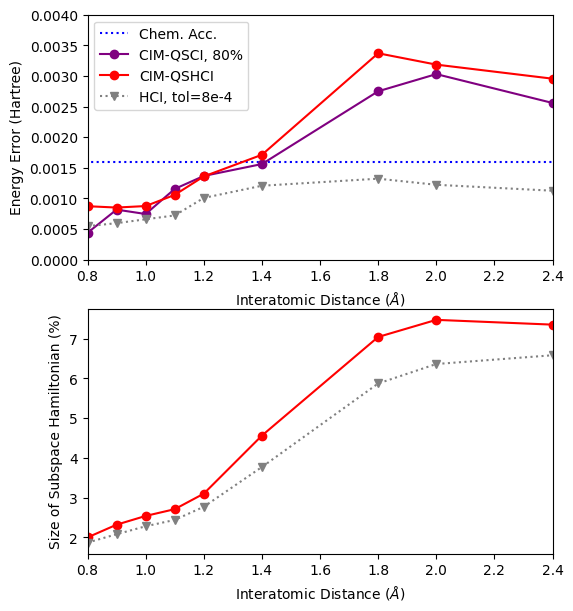}
    \caption{Quantum hardware results for molecular simulation performed using a (10,12) active space. }
    \label{fig:fig2-CIM_QSHCI_N2_Results}
\end{subfigure}

\caption{Absolute error of simulated N$_2$ ground state energy relative to exact diagonalization (upper panel) and resulting sizes of subspace Hamiltonian (lower panel) for different bond lengths. The results of our CIM-QSCI using a subspace Hamiltonian of 80\% are shown as a solid purple line with circles; QSHCI with a variance factor of 1.0 is shown as the red solid line with circles; and QSHCI with a variance factor of 100 is shown as the red dashed line with circles. These results are compared with HCI using tolerances of $5\times10^{-5}$ and $5\times10^{-3}$ (Figure~\ref{fig:fig1-CIM_QSHCI_N2_Results}) and $8\times10^{-4}$ (Figure~\ref{fig:fig2-CIM_QSHCI_N2_Results}). All error curves (top panel) are plotted against the chemical accuracy (0.0016 Hartree), indicated by a blue dotted line. }
\label{fig:CIM_QSHCI_N2_Results}
\end{figure*}

The CIM-QSHCI simulations were performed with variance factors 1.0 and 100.0 for the (10,10) active space and only with 1.0 for the (10,12). For the latter, CIM-QSHCI used the same sampled bitstrings from the quantum hardware measurements as CIM-QSCI. 

The error associated with CIM-QSHCI is very similar to that for CIM-QSCI for all interatomic distances with a subspace Hamiltonian of 80\%. However, the size of the final subspace Hamiltonian is significantly smaller for CIM-QSHCI, showing the advantage of this different post-processing method. The size of the subspace Hamiltonian in CIM-QSHCI ranges from 1.5\% to 10.5\% (Figure~\ref{fig:fig1-CIM_QSHCI_N2_Results}) and 2\% to 8\% (Figure~\ref{fig:fig2-CIM_QSHCI_N2_Results}) of the full size, compared with the 80\% used in CIM-QSCI. This result shows the impact of post-processing on the final result. Furthermore, when the variance factor was set to 100.0 for emulator simulations, the size of the subspace Hamiltonian remained smaller than in CIM-QSCI, and the energy error was lower. This shows that, with the correct variance factor, CIM-QSHCI is a superior algorithm to CIM-QSCI in terms of lower energy error and subspace size.

A comparison between HCI and CIM-QSHCI reveals a slightly different story. In this case, the size of the final subspace Hamiltonians is similar; however, HCI has a lower overall error than CIM-QSHCI. This indicates that although the subspace Hamiltonians include a similar number of determinants, the HCI algorithm identifies more efficient determinants than the CIM-QSHCI. This is likely due to CIM-QSHCI sampling unintended determinants, which was more often the case for samples from the quantum computer than from an emulator. The discrepancies between the emulator and hardware results arise from several factors. First, the quantum circuits used in the hardware experiments were compressed to depths compatible with the hardware, which will have affected the fidelity of the target evolution. Second, our single-bit-flip error mitigation scheme is insufficient to mitigate all types of quantum-device noise.
This makes the CIM-QSHCI sampled-space broader and less efficient than HCI. In contrast to CIM-QSCI, these unwanted states do not affect the quality of the result because QSCI postselects highly frequent states, thereby suppressing spurious states. 

Different tolerance values were used for HCI, which were chosen so that the size of the subspace Hamiltonians was approximately the same as those in CIM-QSHCI. In particular, we observe that for CIM-QSHCI with a variance factor of 1.0 and the HCI set at tolerance $5\times10^{-3}$, this case interestingly resulted in similar energy error for the two methods as shown in the top panel of  Figure~\ref{fig:fig1-CIM_QSHCI_N2_Results}. In contrast, the case of CIM-QSHCI with a variance factor of 100.0 and HCI set to a tolerance of $5\times10^{-5}$ resulted in a clear separation of the two methods, with HCI showing better performance. These results can be explained by the inclusion of the variance factor, which was intended to allow the user to arbitrarily increase the method's accuracy by relaxing the acceptance criteria and thereby increasing the size of the subspace Hamiltonian. Although this works in principle, it is evident that the additional determinants included, using the variance factor, are not the most efficient. This is shown by CIM-QSHCI and HCI performing similarly when the variance factor is set to 1.0, but HCI outperforms CIM-QSHCI when the variance factor is not 1.0. This suggests that there may be a more appropriate way to relax the acceptance criteria, which we leave to future work.

\subsection{CIM-QS(H)CI for simulating Naphthalene}

\begin{figure}[ht]
    \centering
    \includegraphics[width=1.0\linewidth]{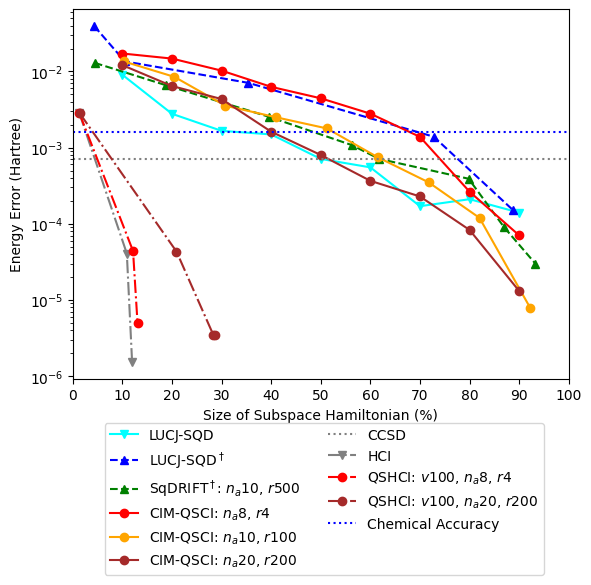}
    \caption{Absolute error of simulated ground state energy of Naphthalene computed using several methods. The CIM-QSCI: red, orange, and brown circles with solid lines, CIM-QSHCI: red and brown circles with dot-dashed lines, using LUCJ-SQD with two LUCJ repetitions: cyan, using LUCJ-SQD with one LUCJ repetition: blue triangles, using SqDRIFT: green triangles. The $n_a$ and $r$ indicate the number of Hamiltonian terms and repetitions used in the evolution. The superscript $\dagger$ indicates data from  \cite{piccinelli_quantum_2025}. The X-axis shows the size of the subspace Hamiltonian as a percentage of the full Hamiltonian size. These quantum algorithms are compared with the classical CCSD and HCI methods. Chemical accuracy (0.0016 Hartree) is shown as the blue dotted line.}
    \label{fig:nap_ham_size}
\end{figure}

In this section, we perform simulation benchmarks with the Naphthalene molecule using model parameters described in Section~\ref{sec:comp_details} and quantum resources summarised in Table~\ref{tab:all_details}. Figure~\ref{fig:nap_ham_size} shows the results of CIM-QS(H)CI for the absolute error of simulating the ground state energy of the molecule relative to exact diagonalization. All the CIM-QS(H)CI simulations were performed on an emulator. These results are compared to several other methods. First, the LUCJ-SQD and SqDRIFT data, taken from \cite{piccinelli_quantum_2025}, which are marked with a $\dagger$ superscript in the legend in Figure~\ref{fig:nap_ham_size}. Second, our implementation of LUCJ-SQD, which uses two LUCJ repetitions. Last, the classical CCSD and HCI methods. We observe several advantages of our CIM-QS(H)CI algorithms compared to LUCJ-SQD and SqDRIFT.  

In general, as expected, increasing the size of the subspace Hamiltonian reduces the error in the CIM-QSCI eigenvalues. In addition, increasing the value of $n_a$ and $r$ also improves the accuracy of the result. Chemical accuracy is achieved when the size of the subspace Hamiltonian is around 70\% for the $n_a$8$r$4 model, 50\% for the $n_a$10$r$100 model, and 40\% for the $n_a$20$r$200 model. These trends are consistent with the results in \cite{piccinelli_quantum_2025} and the simulation of $N_2$ in Sections~\ref{subsec:CIM-QSCI-N2} and \ref{subsec:CIM-QSHCI-N2}. The performance of the $n_a$8$r$4 model is very similar to the LUCJ-SQD with one repetition results of \cite{piccinelli_quantum_2025}, and the $n_a$10$r$100 is similar to SqDRIFT.

The first advantage shown in Figure~\ref{fig:nap_ham_size} is that, by comparing our CIM-QSCI with SqDRIFT for the same model parameters of $n_a$ = 10, we see that SqDRIFT required 5 times as many circuits as CIM-QSCI to achieve a similar accuracy despite CIM-QSCI and SqDRIFT having a similar gate count. The additional circuits required in SqDRIFT may arise from the algorithm including configurations that lack A$_g$ symmetry when selecting the terms used in the modified qDRIFT. As a result, some of the terms used will refer to configurations that do not contribute to the ground state. This leads to redundancies in the quantum circuits, and therefore, more circuits are required to explore enough of the A$_g$ Hilbert space. In the CIM-QSCI algorithm, only A$_g$ CSFs have been used and, as a result, the terms selected in the modified qDRIFT step always relate to configurations that most likely contribute to the ground state. 

The second lower-quantum-resource advantage of our CIM-QSCI lies in the number of qubits required by the algorithm, which is lower than that of the other approaches presented here for all molecules. Third, by comparing the CIM-QSCI with the $n_a$8$r$4 model and LUCJ-SQD with one repetition, two models that achieve similar accuracy (Figure~\ref{fig:nap_ham_size}), the CIM-QSCI yields gate counts significantly lower than LUCJ-SQD. However, this comes at the cost of running four circuits in CIM-QSCI, compared to a single circuit in LUCJ-SQD. This trade-off favours the CIM-QSCI in the NISQ era, given the limited availability of qubit-count resources.

Fourth, by comparing the circuit costs for SqDRIFT \footnote{The gate counts are for SqDRIFT model $n_a$10$r$500 are assumed to be similar to the model $n_a$10$r$1000 reported in \cite{piccinelli_quantum_2025} as the only difference in these models is the number of repetitions.} and CIM-QSCI, it can be seen that the 2-qubit gate count and depth are approximately the same, but that the qubit requirements are lower for CIM-QS(H)CI. This underscores the fact that our CIM-QS(H)CI methods achieve similar or better accuracy with fewer quantum resources. 


Fifth, Figure~\ref{fig:nap_ham_size} shows that the CIM-QSHCI with a variance factor of 100.0 clearly outperforms the CIM-QSCI, LUCJ-SQD, and SqDRIFT, particularly in terms of the size of the subspace Hamiltonian. Two CIM-QSHCI calculations were performed, the first with approximate evolution parameters set as the default values of $n_a$8$r$4 and the second with the parameters set as the larger values of $n_a$20$r$200. This was performed to investigate the impact of the approximate evolution on the final result. For CIM-QSCI, a more accurate evolution improved the overall result of the algorithm. This is also seen for CIM-QSHCI, the more accurate evolution led to more determinants being sampled on the quantum computer, and consequently, the subspace size increased. This resulted in a slightly lower final energy. Comparison with HCI shows that the less accurate approximate evolution performs very similarly, with the final sub-space Hamiltonians being of comparable sizes. The HCI subspace appears to contain more efficient determinants than the CIM-QSHCI subspace, as indicated by the lower overall energy. This was also seen in the other computations of this work. The HCI algorithm can produce an energy estimate with high accuracy in 3 iterations, with a much smaller subspace Hamiltonian ($\approx 11$\%) than the quantum algorithms. This poses a challenge for the quantum algorithms, which will be addressed in future work. 

Finally, a comparison with the classical algorithms: CCSD is also shown in Figure~\ref{fig:nap_ham_size}. All quantum algorithms can outperform CCSD with a sufficiently large subspace Hamiltonian.  
\section{Summary and outlook} \label{sec:conclusion}

This work proposes a new variant of Quantum Select Configuration Interaction (QSCI) \cite{kanno_quantum-selected_2023} based on the Configuration Interaction Matrix (CIM) framework \cite{toloui_quantum_2013, babbush_exponentially_2017} in the first quantization, dubbed CIM-QSCI, and an extension that combines QSCI with Heath-bath CI (HCI), referred to as CIM-QSHCI. These novel CIM-based algorithms include several contributions: 
\begin{itemize}
    \item \textbf{An alternative formulation of the qubit-Hamiltonian} which decomposes the CI-Hamiltonian into a weighted sum of products of Pauli matrices using the fast Walsh-Hadamard transform (FWHT). The benefits of utilising the CIM are the ability to easily remove CSFs that do not have the desired spin-symmetry by considering the molecular point group, a lower memory footprint because the CI-Hamiltonian is constructed in single-precision, and a low qubit requirement of exactly $\lceil \log_2 (N_{CSF})+1 \rceil$. This is consistently lower than similar second-quantized approaches. The latter point makes this approach more appropriate for the NISQ-era devices. 
    \item \textbf{A single bit-flip mitigation scheme} formulated in first quantisation, which reduces the impact that bit-flip errors have on the computation as a post-selection scheme. This comes at a modest cost of one additional qubit, making it quantum resource-efficient.
    \item \textbf{An approximate evolution} using a modified qDRIFT Trotterization \cite{campbell_random_2019}, resulting in relatively short circuits which can be implemented on NISQ devices, which maintain good accuracy.
    \item \textbf{A novel post-processing step} (in CIM-QSHCI) which incorporates elements of the heat-bath CI algorithm and results in significantly smaller sub-space Hamiltonians than traditional QSCI methods. In addition, a user-defined variance parameter allows fine-tuning of the results to increase accuracy while maintaining low computational cost.
\end{itemize}

The CIM-QS(H)CI algorithms were applied to simulate three molecular ground-state problems. First, a (10,12) active space was used for the N$_2$ bond-stretch problem, and the system was simulated on an emulator and on quantum hardware. To reduce the circuit depth to what the hardware can support, an approximate circuit compression was employed with an approximation degree (AD) of 0.50. Further simulations showed that this approximate transpilation could also improve the accuracy of CIM-QSCI by reducing circuit size. Hardware calculations were then performed, showing that the CIM-QSCI error was similar to that of CIM-QSHCI, depending on the size of the subspace Hamiltonians. However, a discrepancy was observed relative to the classical HCI result, which achieved a better overall accuracy with a similarly sized subspace Hamiltonian. This performance difference could be due to accumulated errors in the quantum method which includes the stochastic evolution, circuit compression, and hardware noise. These will be further improved on in future work.

The second ground-state problem was a (10,10) active space, also applied to the bond stretch of N$_2$. This was performed on an emulator with no circuit compression (AD=1.0). These results showed that CIM-QSCI outperformed the classical approach CCSD and the quantum approach LUCJ-SQD. CIM-QSHCI calculations showed that the accuracy and size of the subspace Hamiltonian were similar to those of HCI, depending on the HCI tolerance and on the variance factor in CIM-QSHCI, with a value of 1.0. Increasing the variance factor led to a larger subspace Hamiltonian, thereby improving accuracy. However, HCI outperformed CIM-QSHCI in this case due to a more efficient selection of determinants within its subspace. This poses a challenge to further improve the selection strategy of the determinants in CIM-QSHCI to be competitive to HCI.

Finally, a (10,10) active space with the molecule Naphthalene was investigated. This model was performed on an emulator. It was shown that CIM-QSCI can result in accurate ground state eigenvalues, which have a comparable accuracy to other approaches in the literature, including LUCJ-SQD and SqDRIFT. The accuracy of the result was linked to the accuracy of the approximate evolution performed in the quantum circuit. However, for CIM-QSHCI, improving the accuracy of the approximate evolution yielded a larger subspace Hamiltonian, but only marginally improved the accuracy of the ground-state energy. Using a less accurate approximate evolution led CIM-QSHCI to perform very similarly to HCI.

This work has shown that the novel set of CIM-based algorithms, CIM-QS(H)CI, are able to maintain a high level of accuracy, even when the problem size is increased. In addition, the number of qubits and the two-qubit gate count are consistently low, making this a prime candidate for NISQ devices. The potential for quantum advantage comes in three areas. The first is timing: a speed-up may be possible with this approach compared to classical exact diagonalization. However, the costly overhead for the FWHT makes this more likely for when iterative diagonalization techniques are not used (i.e., denser matrices). The second area of potential advantage is accuracy, especially from CIM-QSHCI. Current results indicate that CIM-QSHCI can perform, at best, just as well as the classical method HCI. Further developments are needed in this space. In this work, the aim was not to surpass classical approaches (such as HCI) but to further extend the capabilities of quantum algorithms, which we have achieved. 

Finally, the third area of quantum advantage, and the most promising, is memory usage. A key benefit of the CIM-QS(H)CI approach is that the full Hamiltonian needs only to be constructed in single (or half) precision. As a result, the memory requirements are much smaller than classical diagonalization techniques. This can allow the user to explore problems on high-performance compute systems that are currently intractable due to memory requirements. It should be noted that this lower precision approach can also be used in other quantum algorithms and is not specific to CIM-QS(H)CI but is an advantage over classical algorithms.

\section*{Supporting information}

Additional information is provided in \cite{graves_supplementary_2026}, including information to re-construct the molecular Hamiltonians used and the raw data used to generate the figures.

\section*{Contributions}
VG proposed the CIM-QS(H)CI methods and performed the numerical simulations. MQ contributed to the complexity analysis of the method and code development for simulations. All authors contributed to the data analysis and interpretation of results. VG drafted this manuscript, and all authors contributed to its final editing. KG provided overall supervision and technical guidance.

\begin{acknowledgments}
We wish to thank the HPC and QCaaS teams at NQCC for support with compute and hardware access.
\end{acknowledgments}

\FloatBarrier

\bibliography{references, QuChem_WP4}
\FloatBarrier
\newpage

\appendix
\section{Mitigation of one-bit flip errors}\label{sec:bit_flip_analysis}

In this section, we justify the bit-flip mitigation scheme presented in the main text. This is done by showing that the inclusion of the bit-flip mitigation scheme has no significant impact when noise is not included in the simulation, but results in a distribution that is closer to the ideal distribution when noise is included. 

Several distributions of sampled basis states are compared from the models:
\begin{itemize}
    \item[\texttt{R0}:] Random sampling.
    \item[\texttt{S1}:] Using a simple binary encoding without the inclusion of noise.
    \item[\texttt{S2}:] Using a simple binary encoding with the inclusion of noise.
    \item[\texttt{B1}:] Using the bit-flip mitigation scheme without noise.
    \item[\texttt{B2}:] Using the bit-flip mitigation scheme with noise.
\end{itemize}
The quantum circuits generated were for the N$_2$ molecule with a bond length of 1.1 \AA~and the (10,10) active space. The terms $n_a$ and $r$ in the approximate evolution were set to the default values of 20 and 10, respectively. The noise model was included using the {\it FakeMarrakesh} backend provided within the Qiskit package.

In this section, we utilise two merits. The first is the cosine similarity, which computes the cosine distance between two vectors. In this work, the two vectors represent the two probability distributions being compared. The cosine similarity, $\delta_{ij}$, is defined as \cite{han_data_2023}:
\begin{align}
    \delta_{ij} = 1 - \frac{P_i \cdot P_j}{||P_i||_2||P_j||_2}
\end{align}
where $P_{i,j}$ are the $i^{\text{th}}$ and $j^{\text{th}}$ probability distributions. The closer $\delta_{ij}$ is to 0, the more similar the distributions are. A $\delta_{ij}$ value of 1 indicates that the distributions are orthogonal and not similar at all.

The second merit is the two-sample Kolmogorov-Smirnov test (KS-test). This merit provides details on whether two distributions are sampled from the same underlying distribution. The Kolmogorov-Smirnov statistic (KS-statistic) is \cite{hodges_significance_1958}:
\begin{align}
    D_{ij} = \max \left| P_i - P_j \right|
\end{align}
where $\max$ is the maximum function. In addition, a p-value can be computed for the KS-Statistic. When the KS-Statistic is large ($\approx$ 1), the p-value will be small indicating that the two distributions are not from the same underlying distribution.

It should be noted that it is difficult to compare two KS-statistics because the extreme values (0 and 1) would not be the same for both statistics, but are instead relative to the distributions inputted. This is in contrast to the cosine similarity, which has well-defined extreme values and, therefore, it's trivial to compare across similarity scores. 

We begin by presenting a comparison with random sampling; \texttt{R0}.

\subsection{Comparison with Random Sampling}

The results from the cosine similarity and KS-tests are shown in Table~\ref{tab:bitmit_random}. Here, a low cosine similarity or KS-statistic score would indicate that the measured distribution is similar to random sampling.
\begin{table}[h]
    \centering
    \caption{Comparisons of distributions \texttt{S1}, \texttt{S2}, \texttt{B1} and \texttt{B2} with random sampling (\texttt{R0}). Scores close to 1.0 indicate that the distributions are not similar to random sampling.}
    \begin{tabular}{c|c | c c}
    \hline\hline
        Distribution & Cosine Similarity & KS-Statistic & p-value \\
    \hline
        \texttt{S1} & 0.82 & 0.99 & 0.00 \\
        \texttt{S2} & 0.58 & 0.97 & 0.00 \\
        \texttt{B1} & 0.82 & 0.99 & 0.00 \\
        \texttt{B2} & 0.62 & 0.97 & 0.00 \\
    \hline\hline
    \end{tabular}
    \label{tab:bitmit_random}
\end{table}
Both metric indicate that the sampled distributions with and without noise are statistically different from random sampling. This is shown most clearly by the KS-statistic which is almost 1.0 in all cases with a p-value of 0.00. This is good because it indicates that the noise from the quantum computer has not resulted in nonsense measurements. In addition, the two noisy distributions (\texttt{S2} and \texttt{B2}) have slightly lower scores than the noiseless ones (\texttt{S1} and \texttt{B1}) due to the presence of noise. The use of the bit-flip mitigation scheme on the noisy measurements (\texttt{B2}) gave a higher cosine similarity score than the non-bit-flip mitigated noisy distribution (\texttt{S2}) showing that the bit-flip mitigation scheme was able to reduce the impact of the noise on the distribution.

\subsection{Comparison with Bit-flip Mitigated Results}

Next, we present a comparison between the distributions that include the bit-flip mitigation scheme and those that do not.
\begin{table}[h]
    \centering
    \caption{Cosine similarity and KS-statistic scores for comparisons between the \texttt{S1}, \texttt{S2}, \texttt{B1} and \texttt{B2} distributions.}
    \begin{tabular}{c |c | c c}
    \hline\hline
        Distributions & Cosine Similarity & KS-Statistic & p-value \\
    \hline
        \texttt{S1} , \texttt{S2} & 0.12 & 0.79 & 0.00 \\
        \texttt{S1} , \texttt{B1} & 0.16 & 0.01 & 0.54 \\
        \texttt{S1} , \texttt{B2} & 0.20 & 0.70 & 0.00 \\
        \texttt{S2} , \texttt{B1} & 0.23 & 0.79 & 0.00 \\
        \texttt{S2} , \texttt{B2} & 0.14 & 0.26 & 0.00 \\
        \texttt{B1} , \texttt{B2} & 0.09 & 0.70 & 0.00 \\
    \hline\hline
    \end{tabular}
    \label{tab:bitmit_stats}
\end{table}
In Table~\ref{tab:bitmit_stats}, the cosine similarity and KS-statistic scores can be seen. The comparison between distributions \texttt{S1} and \texttt{B1} shows that the use of the bit-flip mitigation on a noiseless simulation does not significantly change the measured distribution. This is important because the \texttt{S1} distribution is the ideal distribution and the use of the bit-flip mitigation should not significantly alter it (on an ideal, noiseless simulator). 

By looking at the cosine similarity scores, it can also be seen that the impact of noise on the bit-flip mitigated results is less than on the distributions not using bit-flip mitigation. This is shown by the comparison of distributions \texttt{S1} , \texttt{S2} with \texttt{B1} , \texttt{B2}. The lower cosine similarity score indicates that the bit-flip mitigated results (\texttt{B1} and \texttt{B2}) are more similar than when no bit-flip mitigation is used.

The KS-statistic scores show that the inclusion of noise significantly changes the distribution when no bit-flip mitigation is used (\texttt{S1} , \texttt{S2}). When bit-flip mitigation is used, the similarity between the \texttt{B2} and \texttt{B1} distributions is approximately the same as with the \texttt{S1} distribution. This is because the \texttt{S1} and \texttt{B1} distributions are statistically similar. 

\section{Approximate-transpiliation}\label{sec:approx-transp}

The circuit depth of the (10,12) N$_2$ calculations performed here was too long to be run on hardware via the AWS Braket service. As a result, the depth was reduced using the approximation degree input parameter provided within Qiskit \cite{cross_validating_2019}. This parameter allows the circuit depth to be reduced by decreasing the accuracy of the transpilitation, i.e., making the transpiled circuit similar, but not identical to, the input circuit.

\begin{figure}[ht]
    \centering
    \includegraphics[width=1.0\linewidth]{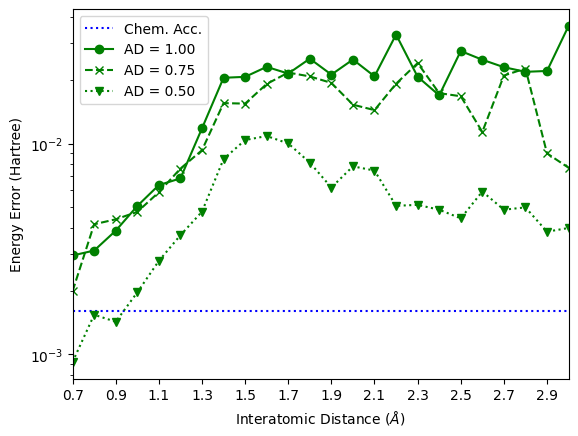}
    \caption{Error of N$_2$ ground state energy for different bond lengths using the (10,10) active space. CIM-QSCI results are shown in green using 60 \% of the subspace Hamiltonian. The degree of the approximate-transpiliation was varied as: 1.00; solid line with circles, 0.75; dashed line with crosses, 0.50; dotted line with triangles. Chemical accuracy (0.0016 Hartree) is shown as a blue dotted line.}
    \label{fig:n2_error_approxtrans}
\end{figure}

To test this approximate-transpiliation, the (10,10) model was emulated using an approximate-degree (AD) of 1.00 (no approximation), 0.75, 0.50, and 0.25. The results are shown in Figure~\ref{fig:n2_error_approxtrans}. Results for the 0.25 AD are not shown as the circuits were reduced to only measurement gates. Also note that when AD = 0.5 there were no 2-qubit gates. In addition, the circuit depth and gate counts are shown in Table~\ref{tab:n2_resources}.
A small change in the AD resulted in the circuits being very similar, and as a result, performing as well as when no approximation is used (when AD = 1.00). However, further reducing the AD to 0.50 resulted in significantly shorter circuits, and therefore a reduction in error-accumulation. This resulted in an improved accuracy of the model in the presence of noise. In addition, the reduction of the circuit depth was sufficient to enable the hardware calculations to be performed. 

As a result, hardware calculations using the (10,12) active space utilised an AD of 0.5. The impact of the AD on the gate count of the (10,12) active space are also shown in Table~\ref{tab:n2_resources}. 

\begin{table}[ht]
\centering
\caption{Impact of approximate transpilation on N$_2$ bond stretch problem with (10,10) and (10,12) active spaces. Number of 2-qubit gate count and depth averaged over all circuits ran for all distances.}
\begin{tabular}{c c ccc}
\hline\hline
Active & Approximation & \multicolumn{3}{c}{Quantum Resources} \\
\cmidrule(lr){3-5}
 Space & Degree & Qubits & Gate & Depth\\
\hline
(10, 10) & 0.50 & 14 & 0 & 0 \\
(10, 10) & 0.75 & 14 & 53 & 40 \\
(10, 10) & 1.00 & 14 & 716 & 596 \\
(10, 12) & 1.00 & 18 & 12,958 & 9,832 \\
(10, 12) & 0.50 & 18 & 43 & 23 \\
\hline\hline
\end{tabular}
\label{tab:n2_resources}
\end{table}

\end{document}